\documentstyle[aps,pre,twocolumn,epsfig]{revtex}
\begin{document}
%
%
\newcommand{\mod}[0]{\mbox{ mod }}
\newcommand{\Abs}[1]{|#1|}
\newcommand{\Tr}[0]{\mbox{Tr}}
\newcommand{\EqRef}[1]{(\ref{eqn:#1})}
\newcommand{\FigRef}[1]{fig.~\ref{fig:#1}}

\title{Computing the diffusion coefficient for intermittent maps:\\
Resummation of stability ordered cycle expansions}

\author{
Carl P. Dettmann\\
Center for Chaos and Turbulence Studies\\
Niels Bohr Institute\\
Blegdamsvej 17, DK-2100 Copenhagen, Denmark\\
and\\
Per Dahlqvist\\ 
Royal Institute of Technology\\
S-100 44 Stockholm, Sweden\\
}
\date{\today}

\maketitle

\begin{abstract}
We compute the diffusion coefficient and the Lyapunov exponent
for a diffusive intermittent map by
means of cycle expansion of dynamical zeta functions.
The asymptotic power law decay of the coefficients of the 
relevant power series are known analytically.
This information is used to resum these power series into
generalized power series around the algebraic branch point
whose immediate vicinity determines the desired quantities.
In particular we consider a realistic situation where all orbits
with instability up to a certain cutoff are known.
This implies that only a few of the power series coefficients are known
exactly and a lot of them are only approximately given.
We develop methods to extract information from these
{\em stability ordered} cycle expansions
and compute accurate values for the diffusion 
coefficient and the Lyapunov exponent.
The method works successfully all the way up to a
phase transition of the map, beyond which the diffusion coefficient and
Lyapunov exponent are both zero.
\end{abstract}

\section{Introduction}

Given the task of computing an average, such as a Lyapunov exponent
or diffusion coefficient of a chaotic system, one can take two different
approaches. 
Firstly, simulation is usually simple and it provides
an answer without bothering to understand the topology
of the flow, but it may suffer from severe convergence
problems.

Secondly, these averages can be extracted from dynamical zeta functions
and their expansions, known as {\em cycle expansions}.
The basic advantage of expanding the average over cycles is that the
asymptotic limit $t\rightarrow \infty$ is already taken from the beginning.
Longer cycles provide corrections to the results obtained 
from shorter ones\cite{DasBuch}.

Real success applying zeta functions has so far only been demonstrated
for quite a restricted class of dynamical systems\cite{DasBuch,AAC,rugh}. 
The topology of the flow should be Markovian --- symbolic
dynamics may be introduced and this symbolic dynamics is of finite
subshift type (meaning that there is only a finite number of 
forbidden substrings). 
In addition the system need to be hyperbolic --- the stability of cycles
is exponentially bounded with length.
The class of systems complying with these two properties is called
Axiom-A. This class is far too restricted to have any major
relevance in applications. 

Success in expanding a zeta function depends on its analytic structure.
Convergence is hampered by singularities close to the zero being studied.
However, if the nature of a disturbing singularity
is known, one can utilize this knowledge in a resummation scheme.
If the singularity is solely due to intermittency
the  convergence problem is thus tamed to a large extent\cite{PDresum}.

To appreciate the relevance of stability ordering of cycle expansions
we imagine a fairly generic system, given by some set of
differential equations.
The problem of finding
periodic orbits in a systematic way
is largely facilitated if one has some symbolic
dynamics. For a few potentials this is possible, for example
the $x^2y^2$ model\cite{PDGR}, the Helium atom\cite{he}, 
the diamagnetic Kepler problem\cite{dia}
and the anisotropic Kepler problem\cite{akp1,akp2}.

For generic flows it is often
not clear what Poincar\'{e} section should be used, and how it should
be partitioned to generate a symbolic dynamics.
Cycles can be detected numerically by
searching a long trajectory for near recurrences.
The long trajectory method for finding cycles 
discussed in\cite{ACEGP,MR} preferentially finds
the least unstable cycles, regardless of their topological length.
If you can find all cycles with stability $\Lambda_p$ less than a certain
cutoff you can use {\em stability} ordered cycle expansions.
Stability ordering was introduced in \cite{PDGR,PDreson}.
It has later been studied more systematically in \cite{DM97,DC}.
It is much easier to implement for
a generic dynamical system than the
curvature expansions which rely on finite subshift approximations
to a given flow.

A general stability ordered cycle expansion looks like 
$\sum_{i=0}^{N_{\rm max}} a_i \exp(-s l_i)$,
where $a_i$ is a monotonically decreasing sequence but $l_i$ is not
monotonic.
In this paper we will restrict our attention to maps.
(It would then be relevant to speak of 
stability truncation rather than stability ordering.)
The expansion looks like 
$\sum_{i=0}^{N_{\rm max}} a_i z^i$ where 
a few of the coefficients may be exact whereas the rest are only approximate.
In particular if the system is intermittent the number of approximate
coefficients greatly exceeds the number of exact ones and the main task of this
paper is to extract the information they carry.
Moreover, we will make use of our a priori knowledge of the power law
decay of the exact coefficients and employ the resummation technique
of ref \cite{PDresum} to improve convergence.

We believe that the idea of stability ordering
has its biggest potential for systems which cannot be described by
a symbolic dynamics of finite subshift type.
But in order to identify the problems due only to intermittency,
we will study a map with complete symbolic dynamics.

\section{Theory}

\subsection{Averages and zeta functions}
\label{s:theory}

A nice introduction to chaotic averages is found 
with proper references in \cite{PCLA}; we will
take a slightly different approach. The reason for this is that the key
step in \cite{PCLA} assumes that the leading zero of a zeta function
is isolated. We will try to avoid this assumption by starting from an
expression for the invariant density in terms of periodic orbits.
The price we pay is that this formula is not rigorously proven for the
intermittent systems we will study.

The aim is to compute averages like
\begin{equation}
\langle  w(x,n)\rangle = \int \rho(x) w(x,n) dz 
\end{equation}
where $\rho(x)$ is the invariant density of the ergodic map 
$x \mapsto f(x)$.

This density can be expressed in terms of periodic orbits via 
\begin{equation}
\rho(x)=\lim_{n \rightarrow \infty} \sum_{p} \sum_{r=1}^{\infty}
\frac{\delta_{n,rn_p}}{|\Lambda_p|^r} \sum_{x_i \in p} \delta(x-x_i)
\end{equation}
where $r$ is the number of repetitions of primitive orbit $p$, having period 
$n_{p}$, and  stability 
$\Lambda_p=\frac{df^{n_p}}{dx}|_{x=x_i}$ with $x_i$ being any
point along $p$. 

The weight $w(x_0,n)$ 
is associated with
the trajectory starting at 
$x_0$ and evolving during $n$ iterations 
in such a way that it is multiplicative
along the flow: $w(x_0,n_1+n_2)=w(x_0,n_1)\cdot w(f^{n_1}(x_0),n_2)$.
As we are dealing with maps, it is simply
$w(x_0,n)=w(x_0,1)\cdot w(f(x_0),1)\cdot w(f^2(x_0),1)\ldots w(f^{n-1},1)$.
The phase space average of $w(x_0,n)$ may now be expanded in terms of periodic
orbits as
\begin{equation}
\lim_{n \rightarrow \infty}
\langle  w(x_0,n)\rangle  = \lim_{n \rightarrow \infty} 
\sum_p n_p \sum_{r=1}^{\infty} w_p^r \frac{\delta_{n,rn_p}}
{\Abs{\Lambda_p^r}}  
 \ \ ,
\label{eqn:tracedef}
\end{equation}
$w_p$ is
the weight along with cycle $p$.
Zeta functions are introduced by observing that the average 
\EqRef{tracedef} may be written as
\begin{equation}
\lim_{n \rightarrow \infty}\langle w(x_0,n)\rangle  = \lim_{n \rightarrow \infty} \frac{1}{2\pi i}
\int_C z^{-n} \frac{d}{dz} \log \zeta_w^{-1}(z) dz
\label{eqn:intlogder}   \ \ , 
\end{equation}
with the zeta function 

\begin{equation}
     1/\zeta_w(z)=\prod_{p}
          \left(1-w_p \frac{z^{n_{p}}}
    {\Abs{\Lambda_{p}} }\right)  \ \ .
\label{eqn:zetaw}
\end{equation}
$C$ is a small contour encircling the origin in clockwise direction.
Eq.~(\ref{eqn:intlogder}) may be verified by inserting the zeta function
(\ref{eqn:zetaw}) and let the integral pick up the residues from $z=0$.
The result can be recast into a sum over residues outside $C$, that is,
it may be related to the analytic structure of the zeta function.

The {\em Lyapunov exponent} can be expressed in terms of a generating
function
\[
\lambda\equiv \lim_{n \rightarrow \infty} \frac{1}{n}
\langle \log |\Lambda(x_0,n)| \rangle 
=\lim_{n \rightarrow \infty}\frac{1}{n}\frac{d}{d\beta}
\langle \Lambda(x_0,n)^\beta\rangle\mid_{\beta=0}
\]

One therefore introduce the multiplicative weight
\begin{equation}
w (x_0,n)=\Lambda(x_0,n )^\beta  \label{eqn:wlam}  \ \ ,
\end{equation}

One can now express the Lyapunov exponent in terms of the associated
zeta function
\begin{equation}
\lambda=
\lim_{n \rightarrow \infty} \frac{1}{n}\frac{1}{2\pi i}
\int_C z^{-n} \frac{d}{d \beta}
\frac{d}{d z}
\log \zeta^{-1}_\lambda(z) \mid_{\beta=0} dz \ \ .
\label{eqn:lamC}
\end{equation}

For a diffusive map $\hat{f}:\bf R \mapsto  R$,
the diffusion coefficient can also expressed in terms of a
generating function
\begin{eqnarray}
D&=&\lim_{n \rightarrow \infty}\frac{1}{2n} 
\langle(\hat{f}^n(\hat{x}_0) -\hat{x}_0)^2\rangle\nonumber\\
&=&\lim_{n \rightarrow \infty}\frac{1}{2n}\frac{d^2}{d\beta^2}
 \langle
e^{\beta (\hat{f}^n(\hat{x}_0)-\hat{x}_0)}\rangle\mid_{\beta=0}
\label{7.5}
\end{eqnarray}
motivating the introduction of the weight
\begin{equation}
w_{D}(x_0,t)=e^{\beta (\hat{f}^n(\hat{x}_0) )-\hat{x}_0)}  \ \ .
\label{eqn:wD}
\end{equation}

If $\hat{f}(\hat{x}+nL)=\hat{f}(\hat{x})+nL$ 
where $\hat{x} \in I$ ($I$ is some interval of length L)
then the map can be reduced to a map $f: I \mapsto I$ on the
{\em elementary cell}.

This may be expressed in terms of a zeta function 
with the weight $w_p$ along cycle $p$ on the elementary
cell given by
\begin{equation}
w_p=e^{\beta \sigma_p}
\end{equation}
where
\begin{equation}
\sigma_p=\sum_{x_i \in p} \left( \hat{f}(x_i)-f(x_i) \right)
\end{equation}
is the corresponding drift in the full system.

The diffusion coefficient may now be expressed in terms of the associated zeta
function

\begin{equation}
D= \lim_{n \rightarrow \infty}\frac{1}{2n} \frac{1}{2\pi i}
\int_C z^{-n} 
\frac{d^2}{d\beta^2}
\frac{d}{d z}
\log \zeta^{-1}_D(z) \mid_{\beta=0} dz   \ \ .
\label{eqn:DC}
\end{equation}
In both \EqRef{lamC} and \EqRef{DC} the asymptotic behavior will determined by 
the leading singularity of the integrand.
Since the integrand is evaluated at $\beta=0$, the singularity
is located at
$z=1$.
If this singularity is isolated the asymptotic result is obtained by simply
integrating around it.
For the intermittent system we are going to consider there
is a complication. 
The singularity is not isolated.
The
zeta function has a branch cut along $\mbox{Re}(z)\geq 1$ and 
$\mbox{Im}(z)=0$.
To extract the asymptotic behavior of these integrals we need to integrate
around this cut.

Let us return to the Lyapunov exponent and assume that
\begin{eqnarray}
1/\zeta_\lambda(z,\beta)&=&[ a_1 (1-z) + O((1-z)^{\gamma}) ]\nonumber\\
& &+ \beta [ b_0  + O(1-z) ] + O(\beta^2)
\label{e:lyapansatz}
\end{eqnarray}
where
$\gamma >1$. This particular assumption will be motivated later in this paper.
We need to evaluate
\begin{eqnarray}
& &\frac{1}{2\pi i}
\int_{\Gamma_0} (1-s)^{-n} \frac{d}{d \beta}
\frac{d}{d s}
\log \zeta^{-1}_\lambda(z(s)) \mid_{\beta=0} ds \nonumber\\
&=&\frac{1}{2\pi i}
\int_{\Gamma_0} (1-s)^{-n} 
\left(-\frac{b_0}{a_1}\right) \frac{1+O(s^{\gamma-1})}{s^2} ds\nonumber\\
&=&-\frac{b_0}{a_1}n+O(n^{2-\gamma})
\end{eqnarray}
where we have changed variable to $s=1-z$.
$\Gamma_0$ is a contour encircling the negative real $s$-axis 
in an anti-clockwise direction.

When evaluating these integrals the following formula is useful
\begin{equation}
\frac{1}{2\pi i} \int_{\Gamma_0} \frac{1}{s^\rho} e^{st}ds=
\frac{t^{\rho -1}}{\Gamma(\rho)}
\end{equation}

The Lyapunov exponent is thus found to be
\begin{equation}
\lambda=-\frac{b_0}{a_1}
\end{equation}

For the diffusion case we assume that
\begin{eqnarray}
1/\zeta_D(z,\beta)&=&[ a_1 (1-z) + O((1-z)^{\gamma}) ]\nonumber\\
& &+ \beta^2 [ c_0  + O(1-z) ] + O(\beta^4)
\label{e:diffansatz}
\end{eqnarray}
We now need to evaluate
\begin{eqnarray}
& &\frac{1}{2\pi i}
\int_{\Gamma_0} (1-s)^{-n} \frac{d^2}{d \beta^2}
\frac{d}{d s}
\log \zeta^{-1}_D(z(s)) \mid_{\beta=0} ds\nonumber\\
 &=&-2\frac{c_0}{a_1}n+O(n^{2-\gamma})
\end{eqnarray}
and
\begin{equation}
D=-\frac{c_0}{a_1}
\end{equation}
To obtain the $a$, $b$ and $c$ coefficients of this section we need to
expand the zeta function~(\ref{eqn:zetaw}) in powers of $z$ around $z=0$
which will be discussed in the next section (\ref{s:expand}) 
Then we will resum
the series around $z=1$ in section~\ref{s:resum}.

\subsection{Expanding zeta functions}
\label{s:expand}

For the Lyapunov exponent calculation we expand the zeta function
\[
1/\zeta_\lambda(z)=\prod_p \left(1-\frac{z^{n_p}}{|\Lambda|^{1-\beta}}\right)
\]
\begin{equation}
=\prod_p \left(1-\frac{z^{n_p}}{|\Lambda|}-
\beta \frac{z^{n_p}\log |\Lambda| }{|\Lambda|}+O(\beta^2)\right)
\end{equation}
\[
=\prod_p \left(1-\frac{z^{n_p}}{|\Lambda|}-
\beta \frac{z^{n_p}\log |\Lambda| }{|\Lambda|}\right)+O(\beta^2)
\]
\[
\equiv \sum_{j=0}^{\infty} \hat{a}_j z^j 
+\beta \left( \sum_{j=0}^{\infty} \hat{b}_j z^j \right)+O(\beta^2)
\]
resulting in two power series.
Similarly for the diffusion calculation we expand
\[
1/\zeta_D(z)=\prod_p \left(1-\frac{z^{n_p}e^{\beta \sigma_p}}{|\Lambda|}\right)
\]
\begin{eqnarray}
1/\zeta_D(z)&=&\prod_p \left(1-\frac{z^{n_p}}{|\Lambda|}
-\beta \frac{z^{n_p}\sigma_p}{|\Lambda|}
-\beta^2 \frac{z^{n_p}\sigma_p^2}{2|\Lambda|}\right.\nonumber\\
& &\left.-\beta^3 \frac{z^{n_p}\sigma_p^3}{6|\Lambda|} \right) +O(\beta^4)
\label{e:evenbeta}
\end{eqnarray}
\[
\equiv \sum_{j=0}^{\infty} \hat{a}_j z^j 
+\beta^2 \left( \sum_{j=0}^{\infty} \hat{c}_j z^j \right)+O(\beta^4)
\]
We restrict our attention to systems with no net drift, that is
\begin{equation}
\lim_{n\rightarrow\infty}\frac{1}{n}\langle
\hat{f}^n(\hat{x}_0)-\hat{x}_0\rangle=0\;\; .
\label{e:nodrift}
\end{equation}
Therefore only even powers of $\beta$ appear in Eq.~(\ref{e:evenbeta}).

The set of coefficients we obtain in this way depends on the truncation
used in the expansion of the infinite product.  For truncation by
topological length, we count cycles up to a given length $N_{\rm top}$.
For maps with a few branches this number is limited to roughly of order
$\sim 10^1$, due to the exponential growth in the number of cycles with the
topological length.
All combinations of cycles with total length less than or equal to
$N_{\rm top}$ are also included, as these contribute to the first
$N_{\rm top}$ coefficients in each series.  Thus we obtain 
$N_{\rm top}$ exact coefficients in each series by topological
length truncation.

For truncation by stability, we count cycles up to a given stability
$\Lambda_{\rm max}$, and combinations where the product of stabilities
is less than $\Lambda_{\rm max}$.  They have
lengths up to $N_{\rm max}$, and so contribute to all of the first 
$N_{\rm max}$
coefficients in each series, but are not the only contributions to
such coefficients. 
They give us an approximation to the zeta function
which for the intermittent case is more accurate than that obtained
from the length truncation, but the values of the coefficients themselves
are not exact beyond some $N_{\rm exact}(\Lambda_{\rm max})$,
a quantity growing logarithmically: $N_{\rm exact}\sim \log \Lambda_{\rm max}$.
For intermittent maps, as the one we will consider,
$N_{\rm max}$ increases as a power of $\Lambda_{\rm max}$
and
$N_{\rm max} \gg N_{\rm exact}$.

Often it is found that stability ordered cycle expansions lead to noisy
results as a function of $\Lambda_{\rm max}$.  This is due to the breaking
of shadowing pairs.  For example a cycle $AB$ usually gives a contribution
roughly equal to and of the opposite sign as the combination of cycles
$A$ and $B$ (we will refer to such a combination as a {\em pseudocycle}).  
This means the total contribution is quite small. The
phenomenon is called shadowing, and is the main mechanism for the rapid
convergence of cycle expansions in hyperbolic systems.  It is still
present to some degree in intermittent systems.  However,
if one such term is included but the other is excluded because they
lie on opposite sides of $\Lambda_{\rm max}$, there may be a substantial
error generated.

Partial shadowing which may be present can be (partially) restored by smoothing
the stability ordered cycle expansions by replacing each term with
inverse pseudocycle stability $\Lambda^{-1}=
(\Lambda_{p_1}\cdots\Lambda_{p_k})^{-1}$ by $S(\Lambda)\Lambda^{-1}$.
Here, $S(\Lambda)$ is a monotonically decreasing function,
with
$S(0)=1$ and
$S(\Lambda > \Lambda_{\rm max})=0$.  

A typical ``shadowing error'' induced by the cutoff is due to two
pseudocycles of stability $\Lambda$ separated by $\Delta\Lambda$, and
whose contribution is of opposite signs.
Ignoring possible weighting factors the magnitude of the resulting term
is of order
$\Lambda^{-1}-(\Lambda+\Delta\Lambda)^{-1}
\approx\Delta\Lambda/\Lambda^2$.
With smoothing there is an
extra term of the form $S'(\Lambda)\Delta\Lambda/\Lambda$, which we
want to minimize.  A reasonable guess might be to keep $S'(\Lambda)/\Lambda$
constant and as small as possible, that is
\[
S(\Lambda)=\left[1-\left(\frac{\Lambda}{\Lambda_{\rm max}}\right)^2\right]
\Theta(\Lambda_{\rm max}-\Lambda)
\]
This function still contains a non-analytic point at
$\Lambda=\Lambda_{\rm max}$, however the discontinuity is now in the
derivative, not in the original function, so a smoothing error estimated by
$S'(\Lambda)/\Lambda$ ($\Lambda<\Lambda_{\rm max}$) is finite.
We use this smoothing function below when evaluating the zeta coefficients,
and demonstrate the improvement numerically.

\subsection{Resumming zeta functions}
\label{s:resum}

The result of the {\em cycle expansions} in sec 2.2
is a set of power series of the form $\sum_i \hat{a}_i z^i$ 
around $z=0$.
And, according to section 2.1, what we need are coefficients from
some kind of (resummed) series around $z=1$.  We now describe a method
of obtaining such a series, along the lines of Ref.~\cite{PDresum}.

Suppose for a moment that the series 
$\sum_{i=0}^{\infty} \hat{a}_i z^i$ has a radius of convergence
exceeding unity.
In a practical calculation we have only a finite number $n$ (say, $N_{\rm top}$
or $N_{\rm max}$) of coefficients $\hat{a}_i$ at our disposal. 
We assume them to be exact, the treatment of the approximate coefficients
from stability ordered expansions are discussed in sec.~\ref{ss:stab}.
Then we can in principle expand it into another truncated (resummed) 
Taylor series around $z=1$.
\begin{equation}
\sum_{i=0}^{n} \hat{a}_i z^i =\sum_{i=0}^{n} a_i (z-1)^i
\end{equation}
This leads to a linear systems of equations which is trivially invertible
\begin{equation}
a_i=\sum_{j=i}^{n} \left(
\begin{array}{c}j\\i\end{array}\right)\; \hat{a}_j
\label{eqn:aninf}
\end{equation}
 In this way one obtains the
standard formulae~\cite{DasBuch}
\begin{eqnarray}
\lambda&=&\frac{\sum(-1)^k\frac{\log\Lambda_1+\ldots+\log\Lambda_k}
{|\Lambda_1\ldots\Lambda_k|}}
{\sum(-1)^k\frac{n_1+\ldots+n_k}{|\Lambda_1\ldots\Lambda_k|}}\\
D&=&\frac{1}{2}\frac{\sum(-1)^k\frac{(\sigma_1+\ldots+\sigma_k)^2}
{|\Lambda_1\ldots\Lambda_k|}}
{\sum(-1)^k\frac{n_1+\ldots+n_k}{|\Lambda_1\ldots\Lambda_k|}}
\label{e:Preddiff}
\end{eqnarray}
where the sums run over all distinct pseudocycles.

This approach is particularly cumbersome for intermittent systems where
$\hat{a}_i$ (as well as $\hat{b}_i$ and $\hat{c}_i$)
decays according to some power law.
Then the coefficients either diverges or converges slowly
as $n\rightarrow \infty$.
So, for intermittent systems the resummed series cannot be a Taylor series, 
it has to be
some generalized power series.

Assume that the asymptotic behavior of the coefficients is a power law
\begin{equation}
\hat{a}_i \sim n^{-(\gamma+1)}
\end{equation}
Then the leading singularity is of the form
$(1-z)^\gamma$, and the simplest possible expansion would be
\begin{eqnarray}
& &\sum_{i=1}^{\infty} a_i (1-z)^i + (1-z)^{\gamma} \sum_{i=0}^{\infty}
\bar{a}_i (1-z)^i\nonumber\\
&=&\sum_{i=0}^\infty \hat{a}_i z^i 
\end{eqnarray}
Having only a finite number $n$ of coefficients $\hat{a}_i$ we propose the
following resummation\cite{PDresum}
\begin{eqnarray}
& &\sum_{i=1}^{n_a} a_i (1-z)^i + (1-z)^{\gamma} \sum_{i=0}^{\bar{n}_a}
\bar{a}_i (1-z)^i\nonumber\\
&=&\sum_{i=0}^n \hat{a}_i z^i +O(z^{n+1})
\end{eqnarray}
If $n_a +\bar{n}_a+2=n+1$ we just get a linear system of equations to solve in
order to to determine the coefficients $a_i$ and $\bar{n}_a$ from 
the coefficients $\hat{a}_i$. 
It also natural to require that $|n_a +\gamma -\bar{n}_a|<1$.

The basic philosophy is to build in as much as information as possible
into the ansatz.
If the original power series correspond to the unweighted zeta function
we know that $a_0=0$. 
The ansatz is thus accordingly modified, we fix $a_0=0$ and modify
$n_a$ or $\bar{n}_a$ so we still get a solvable system of equations.

\begin{figure}
\epsfysize=9cm
\epsfbox{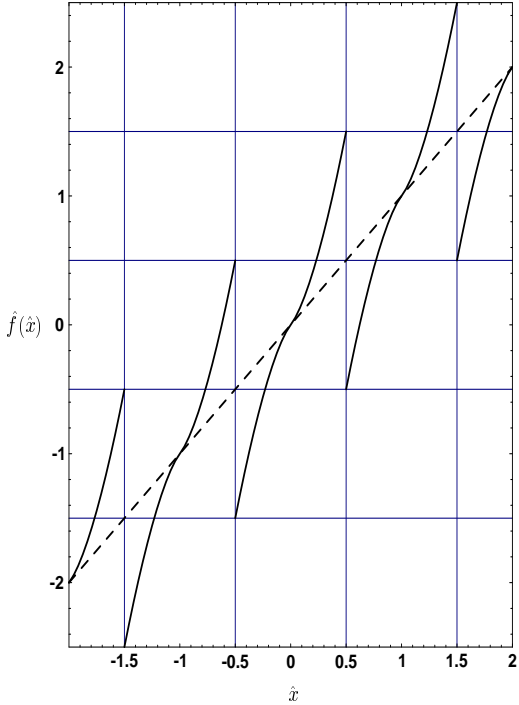}
\caption[]{
The map (\ref{e:map}) for $\alpha=0.7$
}
\label{f:map}
\end{figure}

\section{Numerical studies of an intermittent diffusive map}

\subsection{The map}

In the interval $\hat{x}\in[-1/2,1/2)$, which we call the elementary cell,
our model map, following Ref.~\cite{DC} takes the form
\begin{equation}
\hat{f}(\hat{x})=\hat{x}(1+2|2\hat{x}|^\alpha)\;\;,
\label{e:map}
\end{equation}
The parameter range we consider here is $\alpha\in(0,1)$, where the
Lyapunov exponent
and diffusion coefficient are both nonzero.  For any value of $\alpha$,
this maps the interval $\hat{x}\in[-1/2,1/2)$
monotonically to $[-3/2,3/2)$.  Outside the elementary cell, the map is
defined to have a discrete translational symmetry,
\[
\hat{f}(\hat{x}+n)=\hat{f}(\hat{x})+n\qquad n\in\cal{Z}\;\;.
\]
See Fig.~\ref{f:map}.
A typical initial $\hat{x}$ in the elementary cell diffuses,
wandering over the real line.  The map is parity symmetric,
$
\hat{f}(-\hat{x})=-\hat{f}(\hat{x}),
$
so the average value of $\hat{x}_{n+1}-\hat{x}_n$ is zero,
and there is no mean drift, as expressed in (\ref{e:nodrift}).

We now restrict the dynamics to the elementary cell, that is, we define
\[
x=\hat{x}-[\hat{x}+1/2]\;\;,
\]
where $[z]$ is the greatest integer less than or equal to $z$,
so that   $x$ is restricted to the range $[-1/2,1/2)$.
The reduced map is
\begin{equation}
f(x)=\hat{f}(x)-[\hat{f}(x)+1/2]\;\;.
\label{e:RedMap}
\end{equation}

As discussed in Ref.~\cite{DC}, the intermittency of this map appears
in the form of long cycles near the marginal point with power law
stabilities.  This is in contrast to Axiom-A systems for which
$\Lambda$ may be bounded by exponentials of the topological length.  

The map has three complete branches in the elementary cell.
Symbolic dynamics is introduced by labeling the branches
$\{ -,0,+\}$.
Due to the completeness of  the symbolic dynamics the zeta functions are
approximated by \cite{ACL}
\begin{equation}
1/\zeta_\lambda(z)\approx 
1-\sum_{n=0}^{\infty}\frac{z^{n+1}}
{|\Lambda_{\overline{-0^n}}|^{1-\beta}}
-\sum_{n=0}^{\infty}\frac{z^{n+1}}
{|\Lambda_{\overline{+0^n}}|^{1-\beta}} 
\label{eqn:zeta_appr_lambda}
\end{equation}
and
\begin{equation}
1/\zeta_D(z)\approx 
1-e^{-\beta}\sum_{n=0}^{\infty}\frac{z^{n+1}}
{|\Lambda_{\overline{-0^n}}|}
-e^{+\beta}\sum_{n=0}^{\infty}\frac{z^{n+1}}
{|\Lambda_{\overline{+0^n}}|}
\label{eqn:zeta_appr_diff}
\end{equation}
This approximation may seem crude.
For instance, the zeta functions (\ref{eqn:zeta_appr_lambda},
\ref{eqn:zeta_appr_diff})
fail
to preserve flow conservation.
However in \cite{PDresum} we presented evidence that they 
capture the
leading singularity structure correctly.
This was obtained by comparing coefficients of the piecewise
linear approximation of the intermittent map (sharing the singularity
structure with the approximation above)
by the exact cycle expansion.
The asymptotic behavior of the fundamental cycles is given by
\begin{equation}
\Lambda_{\overline{-0^n}}=
\Lambda_{\overline{+0^n}}
\sim n^{1+1/\alpha}  \ \ , \label{eqn:powerlaw}
\end{equation}
see {\it eg.}\cite{PDresum,DC} for a derivation. 
We obtain immediately from 
(\ref{eqn:zeta_appr_lambda},\ref{eqn:zeta_appr_diff},\ref{eqn:powerlaw})
\begin{eqnarray}
\hat{a}_n&\sim&n^{-1-1/\alpha}\\
\hat{b}_n&\sim&n^{-1-1/\alpha}\log n\\
\hat{c}_n&\sim&n^{-1-1/\alpha}
\end{eqnarray}
This leads to the forms (\ref{e:lyapansatz},\ref{e:diffansatz}) with
$\gamma=1/\alpha$  as long as $\alpha<1$.

For a general orbit we can only bound the stability in the range
\begin{equation}
Cn_p^{1+1/\alpha} < |\Lambda_p| \leq (\mbox{max} |f'|)^{n_p}
=(3+2\alpha)^{n_p}
\end{equation}
so when using stability cutoff we get for the parameters
$N_{\rm max}$ and $N_{\rm exact}$
discussed in section \ref{s:expand}:
\begin{equation}
N{\rm max}\sim \Lambda_{\rm max}^{\frac{\alpha}{1+\alpha}}
\end{equation}
and
\begin{equation}
N_{\rm exact} > \frac{\log \Lambda_{\rm max}}{\log (3+2\alpha)} \ \  .
\end{equation}


\subsection{Resumming topologically ordered cycle expansions}
\label{ss:top}

We will most of the time concentrate on the diffusion coefficient, but
a similar analysis holds for the Lyapunov exponent, to which we
return at the very end.

\begin{figure}
\epsfbox{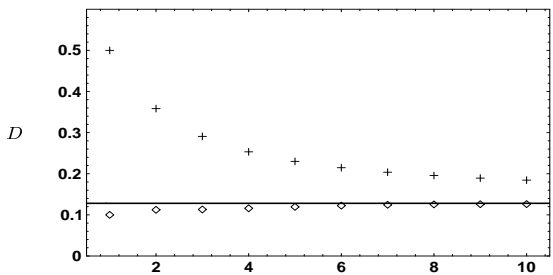}
\caption[]{
The diffusion coefficient at $\alpha=0.7$, from direct simulation (solid
line), topological ordered cycle expansions with (diamonds) and
without (plusses) resummation.
}
\label{f:Dvsn}
\end{figure}

We calculated the diffusion coefficient from resummed cycle expansions
obtained using topological ordering as described in Sect.~\ref{s:resum}
with the number of coefficients $n$ determined by the maximum
topological length, up to $10$.  We also used the direct formula
(\ref{e:Preddiff}), and performed direct simulations with roughly the
same amount of computer time.  The results are shown in Fig.~\ref{f:Dvsn},
showing that the resummation gives much improvement, and is consistent
with direct simulation. 

\subsection{Resumming stability  ordered cycle expansions}
\label{ss:stab}
\begin{figure}
\epsfbox{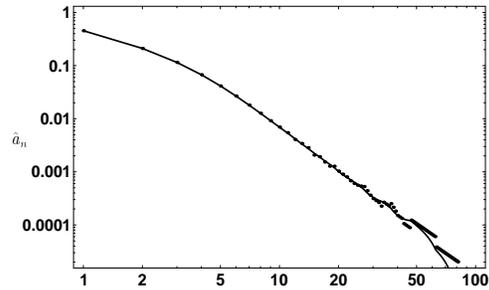}
\caption[]{
The magnitude of the zeta coefficients for stability ordered cycle
expansions at $\alpha=0.7$ unsmoothed (points) and smoothed (solid line).
}
\label{f:smooth}
\end{figure}

Now we come to the central part of our numerical work: the resummation
of stability ordered cycle expansions.  First we calculate the 
$\hat{a}_i$
as described in Sec.~\ref{s:expand}.  The coefficients are all negative
except $\hat{a}_0=1$, and their magnitudes are plotted in
Fig.~\ref{f:smooth}
where we have used $\Lambda_{\rm max} =10^5$ which corresponds to
$N_{\rm exact}=8$ and $N_{\rm max}=81$.
The unsmoothed coefficients are thus exact for 
$n\leq N_{\rm exact}$.
The smoothed are not exact but are still quite accurate.
For $n> N_{\rm exact}$ we clearly see how the unsmoothed begins to
oscillate in an irregular fashion where as the smoothed ones
are stable for much larger $n$. 

\vspace{0.5cm}

The next issue is how to best make use of the information contained in
the $\hat{a}_n$ coefficients.  As pointed out in Sect.~\ref{s:expand}
these coefficients are not exact, but they give a better representation
of the zeta function than the limited number of exact coefficients
obtained from topological ordering.  In order to match the series at
$z=1$, we must
again solve a linear set of equations, but the number of coefficients
($N_{\rm max}$) for intermittent systems is much larger than for the
topological ordering. 
We cannot match such a large number of coefficients in both series, because
the solution would be unstable to the errors in the coefficients, so
we must represent the information contained in the $\hat{a}_n$ in the
(fewer) number of degrees of freedom that the expansion really contains.

There may be more than one solution to this problem; the
solution we use here is to perform {\em two} resummations, the first
from $z=0$ to an intermediate $0<z^{'}<1$, and the second from
$z=z^{'}$ to $z=1$.  
\begin{equation}
\sum_{i=0}^{N_{\rm max}}\hat{a}_iz^i=
\sum_{i=0}^{N_{\rm max}}a^{'}_i(z-z^{'})^i
\end{equation}
which can be explicitly inverted
\begin{equation}
a^{'}_n=\sum_{i=n}^{N_{\rm max}}\left(
\begin{array}{c}i\\n\end{array}\right)\hat{a}_iz^{'(i-n)}
\end{equation}
With $z^{'}$ suitably chosen,
we have thus used the information available in the $\hat{a}_n$ approximately
in proportion to their reliability.  That is, the accurate low order
coefficients appear with large weights in the first few $a^{'}_n$,
while the less accurate high order coefficients appear with small weights.
As we will see, this approach is better than one which simply ignores
the higher order coefficients (this corresponds to putting $z^{'}=0$ below).

As for the topological length truncation, the resummation from $z=z^{'}$ to
$z=1$ leads to a set of linear equations obtained by equating coefficients
in
\begin{eqnarray}
& &\sum_{i=1}^{n_a} a_i (1-z)^i + (1-z)^{\gamma}
\sum_{i=0}^{\bar{n}_a}\bar{a}_i (1-z)^i\nonumber\\
&=&\sum_{i=0}^{n^{'}}a^{'}_i(z-z^{'})^i +O(z^{n+1})
\end{eqnarray}
Again, we adjust $n_a$ and $\bar{n}_a$ so as to obtain a consistent series
in powers of $z-1$ and a consistent set of linear equations.

\begin{figure}
\epsfysize=10cm
\epsfbox{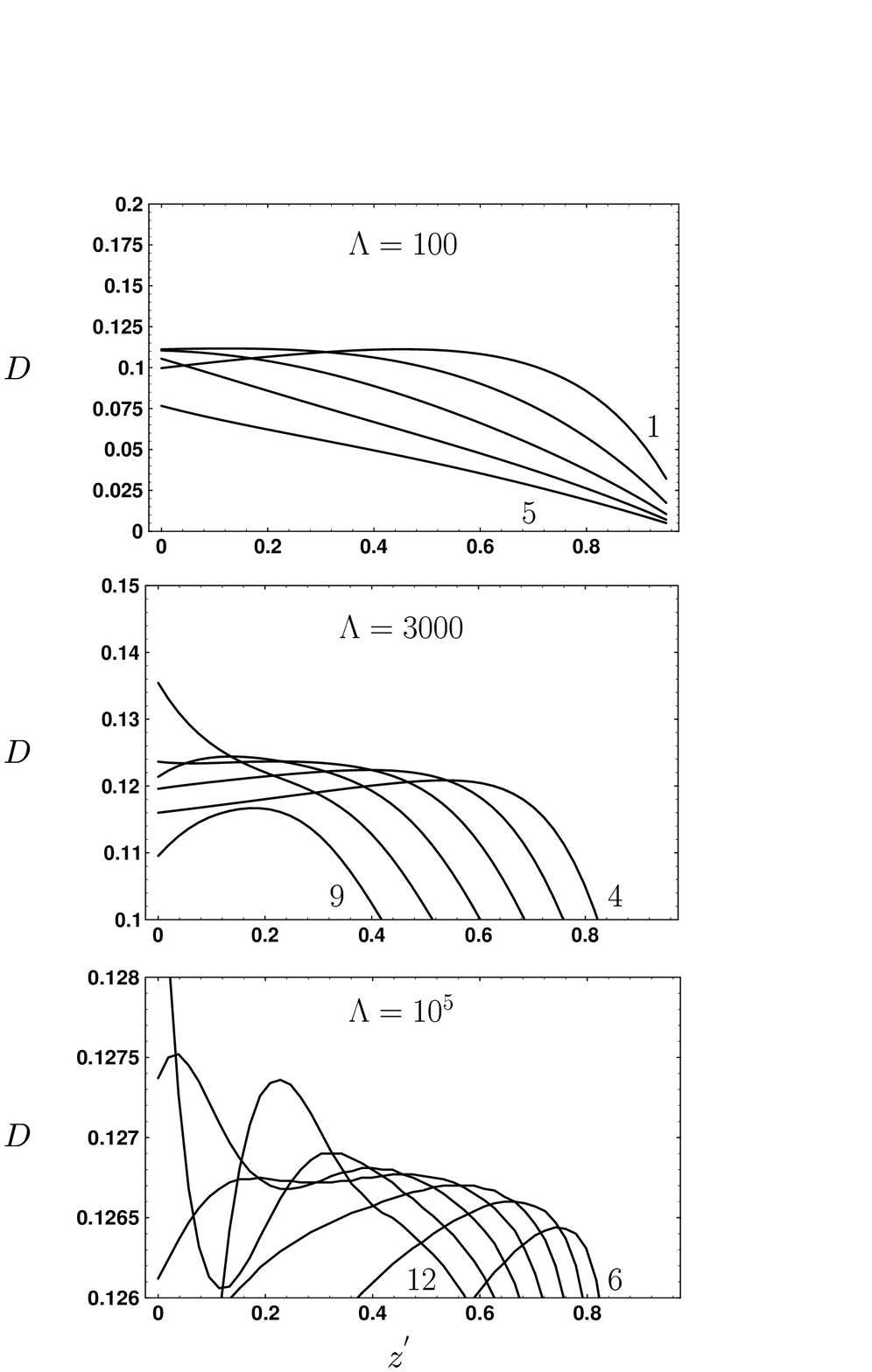}
\caption[]{
Diffusion coefficient calculated using resummed stability ordered cycle
expansions, showing dependence on $n^{'}$ (labels on curves) and
$z^{'}$ (horizontal axis).  Note the scales on the vertical axes.
}
\label{f:var}
\end{figure}

\begin{figure}
\epsfbox{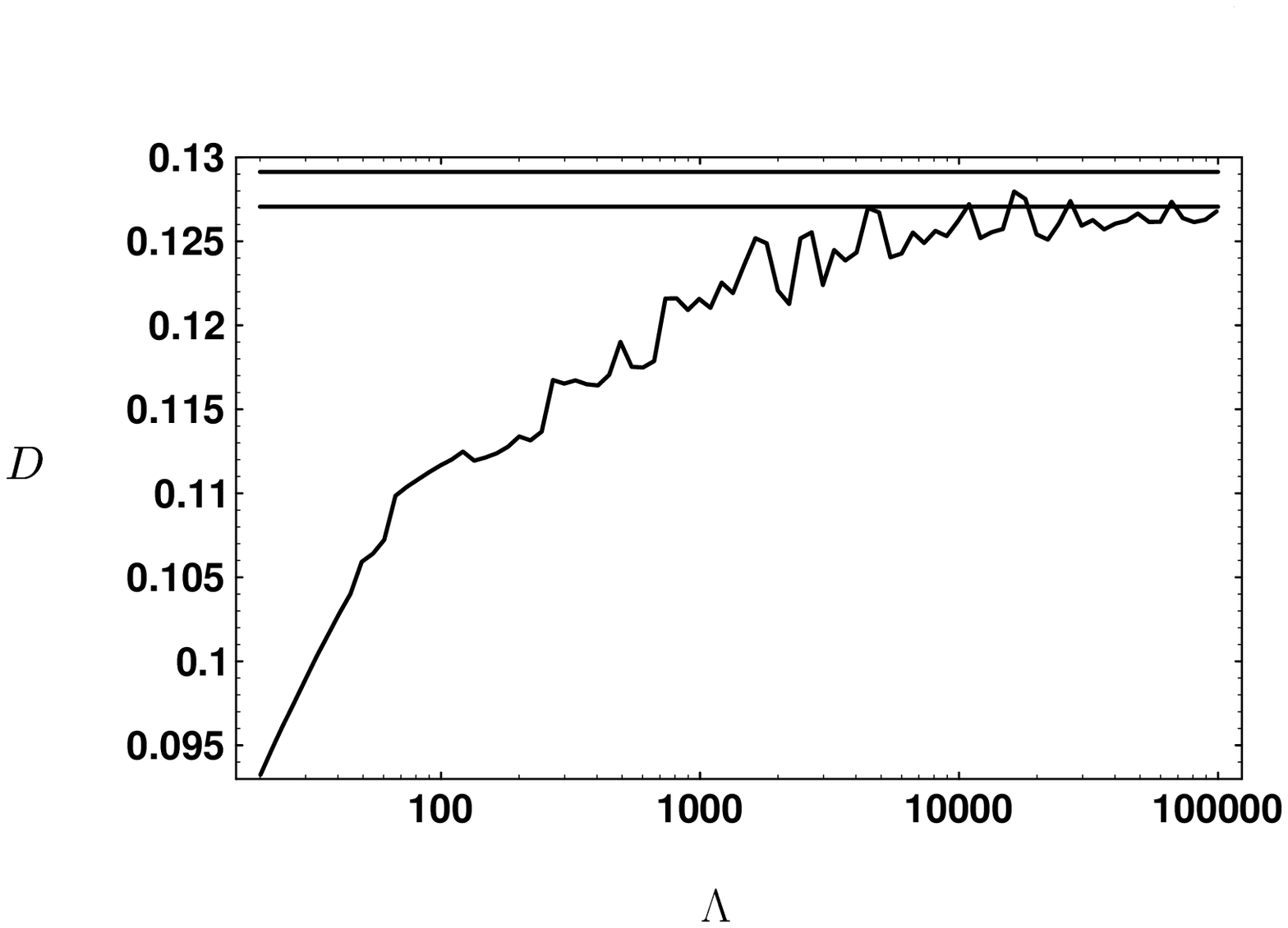}
\caption[]{
Diffusion coefficient calculated using resummed stability ordered cycle
expansions, showing dependence on the stability cutoff.  The horizontal
lines are the range indicated by direct simulation (see also discussion
in text).
}
\label{f:stabresum}
\end{figure}

We have two parameters in the double resummation scheme, $n^{'}$ and $z^{'}$.
The idea is to calculate $\lambda$ or $D$ for a range of both parameters
and look for the most consistent solution.  The general behavior is shown
in Fig.~\ref{f:var}.  For each stability cutoff $\Lambda_{\rm max}$, small values of
$n^{'}$ lead to a variation of $D$ with $z^{'}$ which has a single maximum.
Larger values of $n^{'}$ lead to functions that are either monotonically
decreasing or oscillatory.  We estimate the diffusion coefficient by
finding the maximum for the largest value of $n^{'}$ before monotonic or
oscillatory behavior sets in.  The convergence of this method with the
stability cutoff is shown in Fig.~\ref{f:stabresum}.

This figure also contains the direct simulation results, obtained by
estimating the left side of Eq.~(\ref{7.5}) for $3\times 10^3$ iterations
over a sample of $3\times 10^3$ trajectories, similar to the computer
time required to find the cycles with $\Lambda<10^5$.
The errors were obtained by looking at the scatter in this statistical
sample of trajectories; for intermittent maps the diffusion coefficient
always tends to be too high because long intermittent episodes are not
sampled sufficiently.  Close to the phase transition at $\alpha=1$ convergence
is practically logarithmic in the number of iterations, with exponentially long
times required to achieve convergence.  For example, with the numerical
procedure described above we find
$D=0.0524\pm0.0005$ at $\alpha=1$ where we know $D=0$.
Even at $\alpha=0.7$, a reasonable
distance from the transition, the resummed cycle expansion result is more
accurate than direct simulation.

Our final value for the diffusion coefficient at $\alpha=0.7$ is
$D=0.1267\pm0.0003$ with the resummation method.  
It is then quite compatible with the topological ordering discussed in
Sect~\ref{ss:top}, which yielded the result $D=1.262\pm0.0003$.
In this example, it is clear that resummed cycle expansions, whether
ordered by topological length or stability provide an accurate method
of analyzing intermittent systems.  Stability ordering is most important
in more complicated systems where topological ordering is not a realistic
alternative.  There is an additional advantage with the stability ordered
expansion in that it provides a large number of approximate coefficients,
thus facilitating a numerical estimate of the power law if it is not
known analytically. Recall that
this power law is  used in the resummation ansatz,
and was absolutely essential for the good result in Sect~\ref{ss:top}.

\subsection{The phase transition}

\begin{figure}
\epsfbox{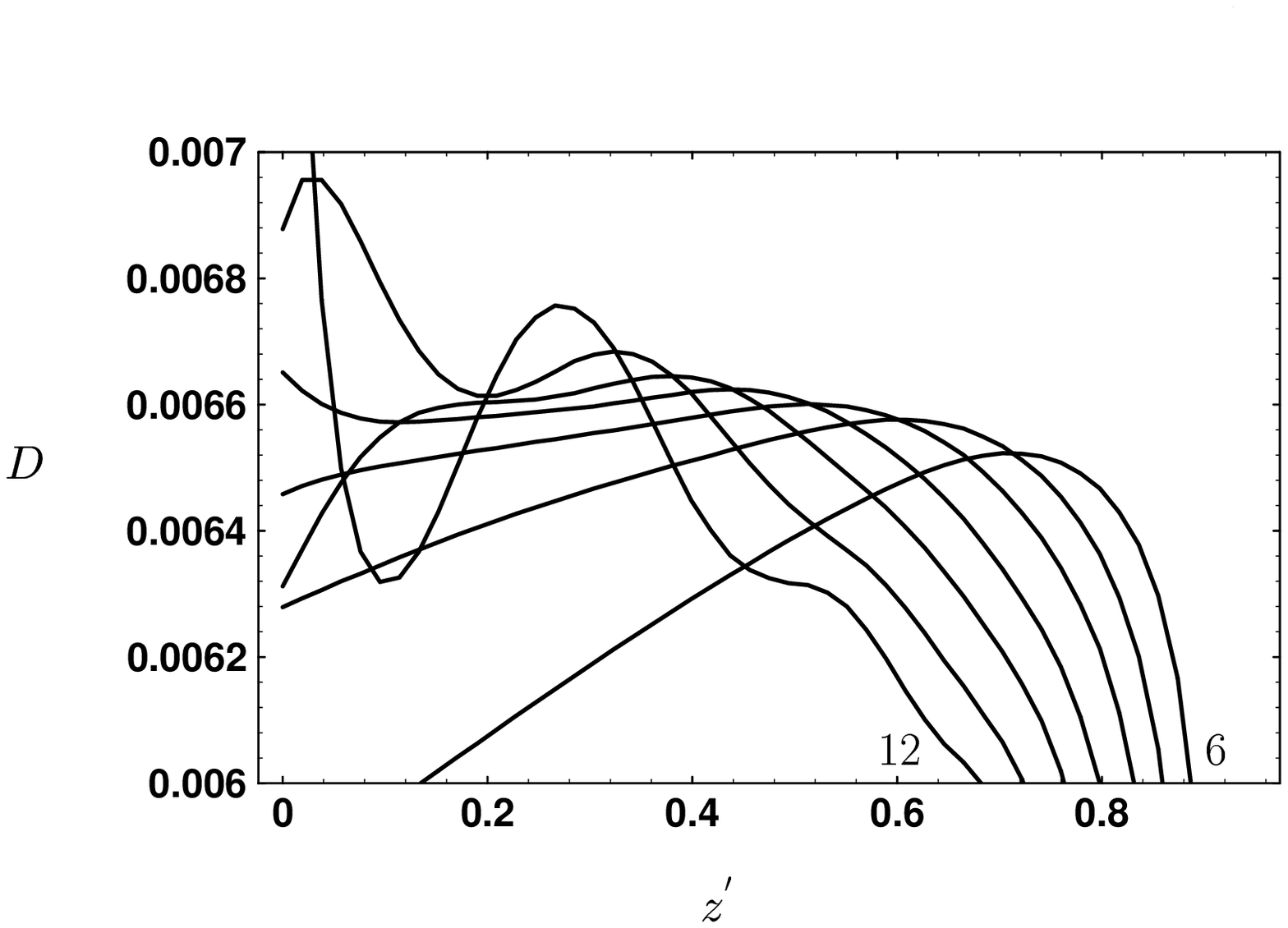}
\caption[]{
Diffusion coefficient calculated using resummed stability ordered cycle
expansions for $\alpha=0.99$.
}
\label{f:0.99}
\end{figure}

\begin{figure}
\epsfbox{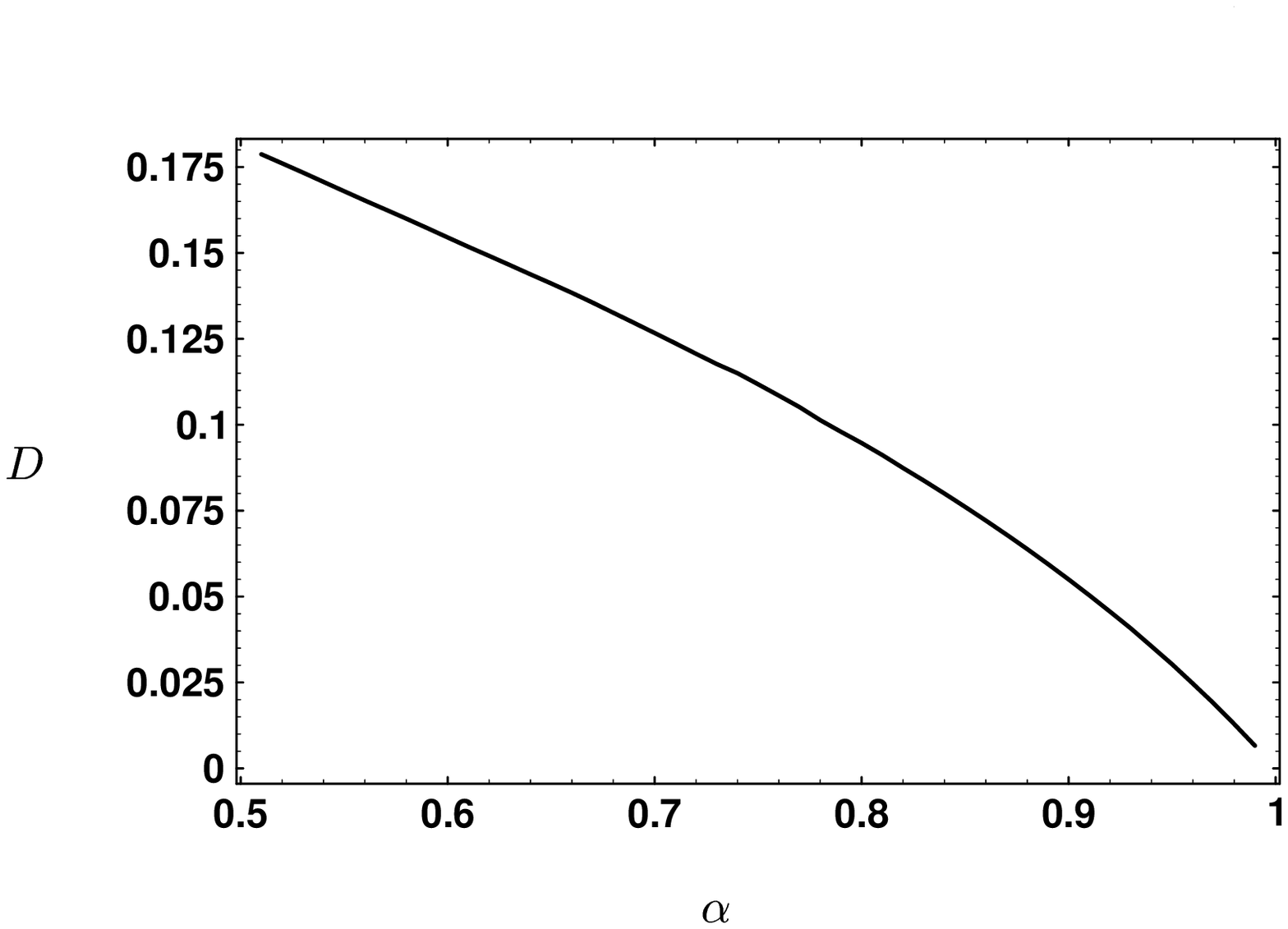}
\caption[]{
The diffusion coefficient as a function of the parameter $\alpha$, showing
the approach to the phase transition at $\alpha=1$, beyond which $D=0$.
}
\label{f:phasediff}
\end{figure}

Having gained confidence in the resummation method for $\alpha=0.7$, far
from the phase transition at $\alpha=1$, we now vary $\alpha$, including
values for which direct simulation is totally impractical, due to
logarithmically slow convergence.  At $\alpha=0.99$ we obtain
Fig.~\ref{f:0.99} for the
diffusion coefficient, showing a consistent value of $D=0.0066\pm0.0001$.
Plotting $D$ vs $\alpha$ (Fig.~\ref{f:phasediff}) we find a linear
dependence near the phase transition at $\alpha=1$.

\begin{figure}
\epsfbox{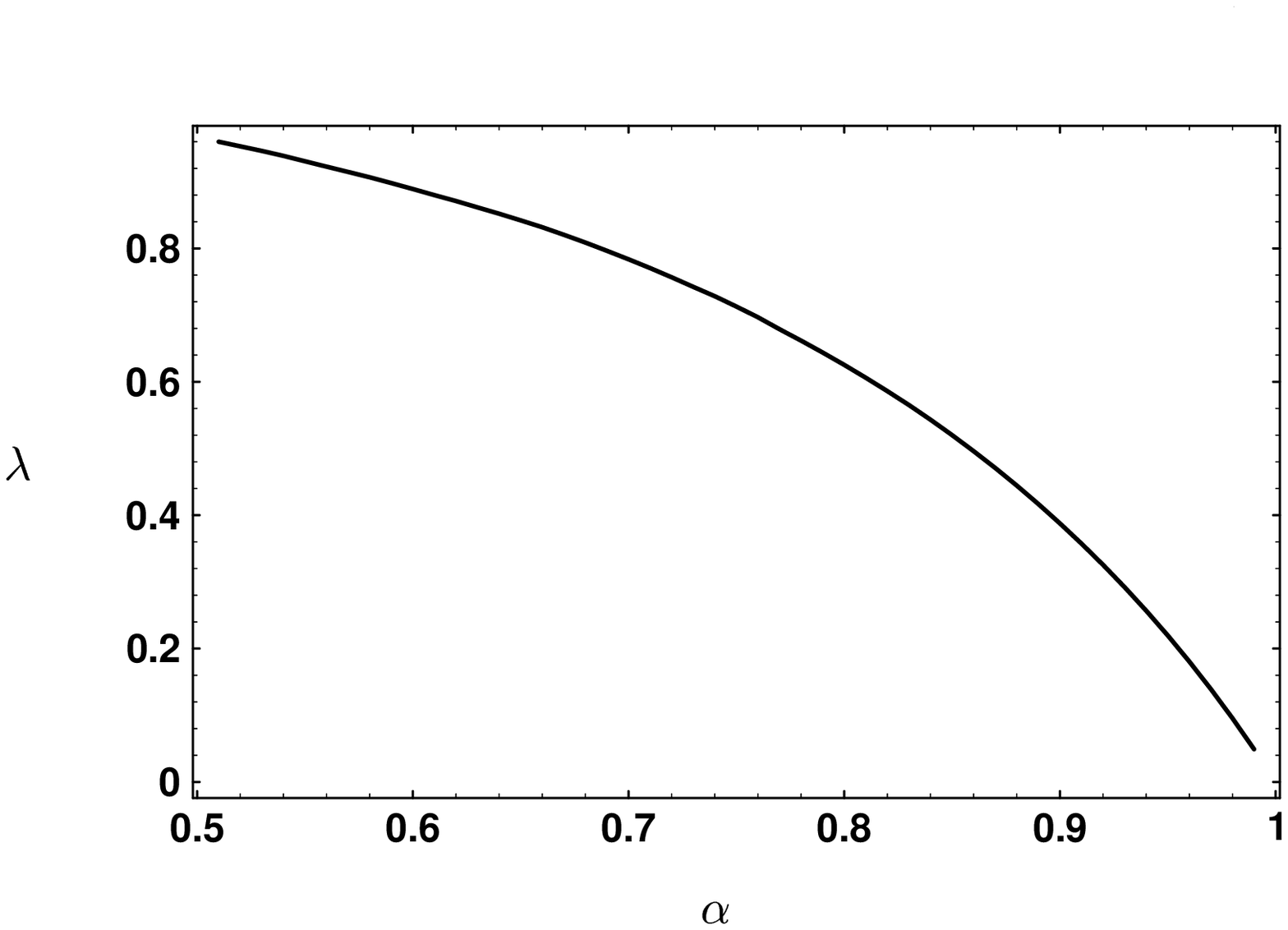}
\caption[]{
The Lyapunov exponent as a function of the parameter $\alpha$, showing
the approach to the phase transition at $\alpha=1$, beyond which $D=0$.
}
\label{f:phaselyap}
\end{figure}

Finally we performed the same analysis for the Lyapunov exponent, which
has a similar dependence on $\alpha$, shown in Fig.~\ref{f:phaselyap}

\section{Conclusion}
We have demonstrated that resummed stability ordered cycle expansions can
provide accurate estimates of dynamical averages for intermittent maps,
even close to a phase transition.  This analysis could equally apply to
maps with uncontrolled symbolic dynamics, as long as a reliable method
exists for locating the cycles.

Our methods can also be applied without
much modification to flows.
Then the variable $z$ is  replaced by $\exp(-s)$ and the cycle
expansion is actually
a Dirichlet
series, $\sum_i a_i \exp(-sl_i)$, where  the lengths of the pseudo
orbits $l_i$ are not restricted to integer values. 
With an additional
resummation step at $s'$ (corresponding to $z'$ in this paper), 
the zeta function may be represented
as a standard power series, thus allowing it to be matched to a 
generalized power
series
at $s=0$. 
For intermittent systems $s=0$ is again a branchpoint, and information
about it can be obtained  from the methods described in
\cite{PDreson,PDsin}
or numerically from the stability ordered expansion.



\vspace{1cm}

C.~D.~ is supported by the Danish Research Academy.
P.~D.~ is supported by the Swedish Natural Science
Research Council (NFR) under contract no.
F-AA/FU 06420-311.
We thank the G\"{o}ran Gustafsson foundation and
NORDITA for support.


\end{document}